\documentclass[preprint,onecolumn,showpacs,floatfix,preprintnumbers,amsmath,amssymb,superscriptaddress]{revtex4-1}
\usepackage{graphicx}
\usepackage{dcolumn}
\usepackage{bm}
\usepackage{color}
\usepackage[latin1]{inputenc}
\usepackage[urlcolor=blue]{hyperref}
\hypersetup{backref, colorlinks=true}

\begin{document}


\title{Thermopower evolution in Yb(Rh$_{1-x}$Co$_x$)$_2$Si$_2$}

\author{U.~Stockert}\email{ulrike.stockert@cpfs.mpg.de}\affiliation{MPI for Chemical Physics of Solids, D-01187 Dresden, Germany}
\author{C. Klingner}\affiliation{MPI for Chemical Physics of Solids, D-01187 Dresden, Germany}
\author{C. Krellner}\affiliation{MPI for Chemical Physics of Solids, D-01187 Dresden, Germany}\affiliation{Physikalisches Institut, Johann Wolfgang Goethe-Universit\"{a}t, D-60438 Frankfurt am Main, Germany}
\author{V. Zlati\'{c}}\affiliation{Department of Physics, Faculty of Science, University of Split, HE-21000 Split, Croatia}
\author{C. Geibel}\affiliation{MPI for Chemical Physics of Solids, D-01187 Dresden, Germany}
\author{F.~Steglich}\affiliation{MPI for Chemical Physics of Solids, D-01187 Dresden, Germany}\affiliation{Center for Correlated Matter, Zhejiang University, Hangzhou, Zhejiang 310058, China}\affiliation{Institute of Physics, Chinese Academy of Science, Beijing 100190, China}

\date{\today}

\begin{abstract}
We present thermopower measurements on Yb(Rh$_{1-x}$Co$_x$)$_2$Si$_2$. Upon cobalt substitution the Kondo temperature is decreasing and the single large thermopower minimum observed for YbRh$_2$Si$_2$ splits into two minima. Simultaneously, the absolute thermopower values are strongly reduced due to a weaker exchange coupling between the $4f$ and the conduction electron states with increasing $x$. Pure YbCo$_2$Si$_2$ is considered a stable, trivalent system. Nevertheless, we still observe two minima in the thermopower indicative of weak residual Kondo scattering. This is in line with results from photo emission spectroscopy revealing a tiny contribution from Yb$^{2+}$. The value at the high-$T$ minimum in $S(T)$ is found to be proportional to the Sommerfeld coefficient for the whole series. This unexpected finding is discussed in relation to recent measurements of the valence and Fermi surface evolution with temperature.  
\end{abstract}


\maketitle

\section{Introduction}

Ce and Yb-based heavy fermion (HF) systems usually exhibit large absolute values of the thermopower $S$ and a characteristic temperature dependence $S(T)$ reflecting Kondo interaction and crystal electric field (CEF) splitting of the $4f$ multiplet. In case of Ce (Yb) the thermopower below room temperature is positive (negative) for the most part with two maxima (minima), one around the Kondo temperature $T_{\mathrm{K}}$ due to scattering on the ground state doublet~\cite{TB-87-1,TB-97-1,TB-05-3} and one at higher $T$ due to combined scattering on the ground state and thermally populated CEF levels. For a CEF level at $k_{\mathrm{B}}T_{\mathrm{CEF}}$ above the ground state this second extremum is expected at about 0.3-0.6~$T_{\mathrm{CEF}}$~\cite{TB-05-3,TB-76-1,TB-86-1}. A single CEF extremum is usually found also in case of more than one excited CEF level due to the fact that thermal population of excited levels takes place over a considerable temperature range. 

Chemical substitution is widely used to induce chemical pressure and to tune the ground-state properties of HF compounds, e.g., with the aim of driving the system to a quantum critical point (QCP) or an intermediate valence (IV) state. This is reached by changing the hybridization strength and the exchange coupling between the $4f$ and conduction electrons and consequently the magnetic coupling constant $J$ upon substitution. The concomitant change of energy scales is reflected in the thermopower mainly as a shift of the maxima or minima as observed, e.g., in Ce$_{1-x}$La$_x$Ni$_2$Ge$_2$~\cite{US12a}. If the Kondo and CEF energy scales get close, the two features may even merge into a single one, e.g., for Ce(Ni$_{1-x}$Pd$_x$)$_2$Si$_2$~\cite{TB-01-3} or Yb(Ni$_x$Cu$_{1-x}$)$_2$Si$_2$~\cite{TB-99-3}. On the other hand, if $J$ is strongly reduced, it is expected that the large thermopower values are reduced and the extrema eventually disappear. In this paper we present an example for this situation, namely Yb(Rh$_{1-x}$Co$_x$)$_2$Si$_2$. Cobalt substitution on the rhodium site leads to a very strong reduction of the exchange coupling thus driving the system from a HF state in YbRh$_2$Si$_2$ to a stable trivalent state in YbCo$_2$Si$_2$.

YbRh$_2$Si$_2$ is a well-known and widely studied HF system, which exhibits non-Fermi-liquid behavior close to an antiferromagnetic QCP~\cite{YRS-00-1,YRS-02-2,YRS-03-2}. It has a ground state Kondo scale $T_{\mathrm{K}}$ of about 25-30~K~\cite{YRS-06-8,UK08a} with a mean Yb valence of 2.93~\cite{YRS-11-1} at low $T$ and higher CEF doublets at 200-290-500~K, respectively~\cite{YRS-06-4}. The other end member of the series, YbCo$_2$Si$_2$, has very week Kondo interaction with a characteristic temperature of less or about 2~K~\cite{YRS-11-5,YRS-11-4} and excited CEF doublets at 46-150-350~K~\cite{Gor00}. Thus, both the Kondo and the CEF scales are significantly lower than in YbRh$_2$Si$_2$. Generally, YbCo$_2$Si$_2$ is considered a stable, trivalent system. However, various experimental probes revealed a tiny residual hybridization between $4f$ and conduction electron states, such as photoemission spectroscopy (PES)~\cite{YRS-11-4}, resonant X-ray emission spectroscopy, and angle-resolved PES (ARPES)~\cite{YRS-14-1}. Likewise, the Kondo-type increase in the electrical resistivity $\rho (T)$ towards low $T$ and the slightly enhanced Sommerfeld coefficient $\gamma _0 = 0.13$~J mol$^{-1}$ K$^{-2}$ point to a small Yb$^{2+}$ contribution to the ground state~\cite{YRS-11-5}. 

The full substitution series Yb(Rh$_{1-x}$Co$_x$)$_2$Si$_2$ has been studied in detail in Ref.~\cite{YRS-11-4}. It exhibits a complex magnetic phase diagram with different magnetic phases below 2~K~\cite{YRS-11-4,YRS-08-3,YRS-13-1}. However, in our investigation we focus on the temperature region above 2~K. Substitution of Rh by Co has two major effects~\cite{YRS-11-4}: (1) With increasing Co content the exchange coupling between $4f$ and conduction electrons is strongly reduced. This was demonstrated by PES revealing a gradual lowering of the Yb$^{2+}$ intensity with increasing $x$. It is confirmed by the concomitant lowering of the Sommerfeld coefficient $\gamma_0$. (2) Simultaneously, the Kondo temperature is reduced. This is evidenced by an increase of the low-T entropy and the shift of the maximum in $\rho (T)$ towards lower $T$. In fact, above a Co concentration of about 50~\% the Kondo interaction is no longer the most relevant exchange interaction. Instead, RKKY interaction and the magnetic ordering of almost unscreened Yb$^{3+}$ moments dominate the low-$T$ properties. Following the notation in Ref.~\cite{YRS-11-4} we, therefore, use the term $T_{4f}$ instead of $T_K$ for the entropy-derived characteristic temperature.
  
In this paper we present thermopower measurements on Yb(Rh$_{1-x}$Co$_x$)$_2$Si$_2$. As expected, the lowering of $T_K$ with increasing Co concentration leads to the appearance of a low-$T$ minimum in the thermopower, similar to the case of Lu substitution on the Yb place~\cite{UK08a}. Simultaneously, the absolute thermopower values are rapidly reduced due to the lowering of the exchange coupling between $4f$ and conduction electrons. The value at the high-$T$ minimum is found to be proportional to the Sommerfeld coefficient. For YbCo$_2$Si$_2$ we still observe the characteristic temperature dependence with two minima in $S(T)$ as a result of a weak residual hybridization between $4f$ and conduction electron states.


\section{Experimental details}

We investigated single crystals of Yb(Rh$_{1-x}$Co$_x$)$_2$Si$_2$ with $0 \leq x \leq 1$ grown by an indium-flux technique. Samples with $x > 0$ stem from the same batches as those characterized in Ref.~\cite{YRS-11-4}. The Co content $x$ determined from energy-dispersive x-ray diffraction has a relative error of less than 1 at.~\%~\cite{YRS-11-4}. Data for pure YbRh$_2$Si$_2$ ($x = 0$) were taken from Ref.~\cite{UK08a}. 

The thermopower was measured between 2~K and 300~K using the thermal transport option of a PPMS. It applies a relaxation-time method with a low-frequency square-wave heat pulse generated by a resistive heater. Two Cernox sensors are utilized for measurement of the temperature difference along the sample. In all measurements the heat current was applied within the $ab$ plane of the plate-like crystals.


\section{Results}

\begin{figure}
\begin{center}
\includegraphics[width=0.7\textwidth]{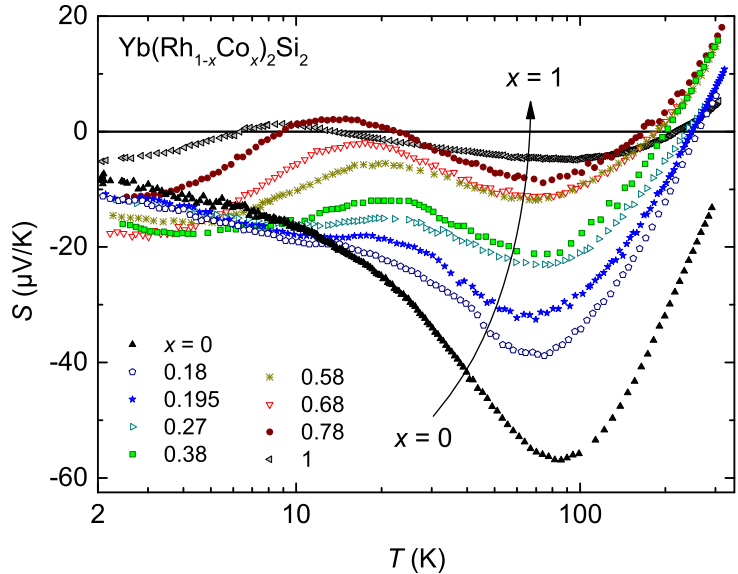}
\end{center}
\caption{(Color online) Thermopower of Yb(Rh$_{1-x}$Co$_x$)$_2$Si$_2$. 
Data for YbRh$_2$Si$_2$ ($x = 0$) have 
been published previously in Ref.~\cite{UK08a}. \label{SvsT}}
\end{figure}

The thermopower $S(T)$ of all investigated samples of Yb(Rh$_{1-x}$Co$_x$)$_2$Si$_2$ is shown in Fig.~\ref{SvsT}. Pure YbRh$_2$Si$_2$ exhibits a large negative thermopower with a single minimum around 80~K that was ascribed to combined Kondo scattering from the ground state and higher CEF levels~\cite{UK08a,YRS-06-5}. Substitution of Rh by Co leads to a lowering of the absolute thermopower values around that minimum, while the minimum itself and its position are rather stable. A second minimum evolves, that shifts to lower $T$ with increasing Co content. For $x \geq 0.68$ this minimum lies around or below 2 K. Its existence is confirmed by the fact that the thermopower must reach zero in the zero-temperature limit. At highest temperatures a crossover to positive thermopower values is observed for samples with cobalt substitution. A positive thermopower is also found at intermediate temperatures for $x \geq 0.78$. These observations are probably due to a positive diffusion contribution to the thermopower from normal (light) charge carriers. In fact, the nonmagnetic reference of YbRh$_2$Si$_2$, LuRh$_2$Si$_2$, exhibits a small positive thermopower~\cite{UK08a}.

The overall behavior of $S(T)$ of Yb(Rh$_{1-x}$Co$_x$)$_2$Si$_2$ can be most simply explained by a lowering of the ground state Kondo scale $T_\mathrm{K}$ accompanied by a weakening of the Kondo interaction upon substitution of Rh by Co and an almost constant characteristic temperature for the CEF level splitting. In this picture, the increasing separation of the Kondo and CEF energy scales gives rise to the splitting of the single large minimum for YbRh$_2$Si$_2$ into two minima for $x \geq 0.195$. With further increasing Co concentration the low-$T$ minimum shifts to lower $T$ as $T_\mathrm{K}$ is decreasing. The concomitant weakening of the Kondo interaction leads to the lowering of the absolute thermopower values, especially around the high-temperature minimum. 

\begin{figure}[tb]
\begin{center}
\includegraphics[width=0.7\textwidth]{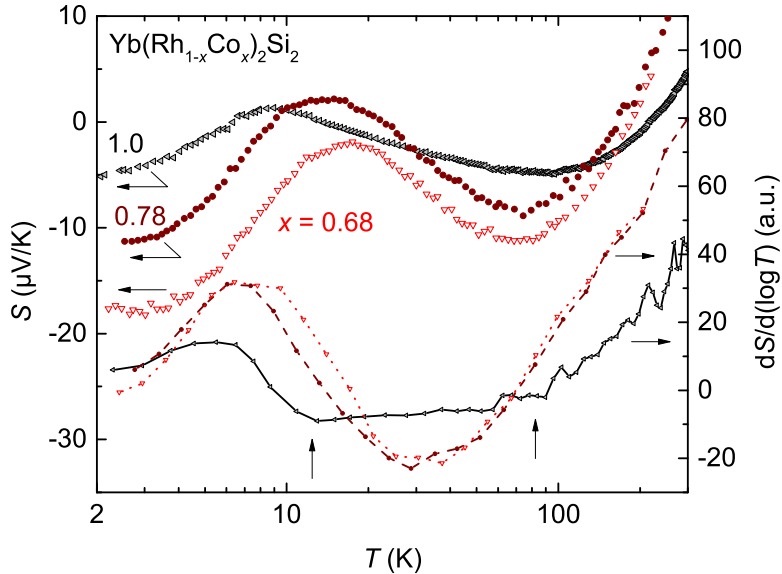}
\end{center}
\caption{(Color online) Thermopower (left axis) and derivative of the thermopower (right axis) on a logarithmic $T$ scale of Yb(Rh$_{1-x}$Co$_x$)$_2$Si$_2$ for $x = 0.68$, 0.78, and 1. The minimum in $S(T)$ is much broader for pure YbCo$_2$Si$_2$ than for the two samples with mixed Co/Rh content due to a low-lying CEF level. This differing behavior is also found in the derivatives of the thermopower on a logarithmic $T$ scale. A clear minimum around 30 K is seen for $x = 0.68$ and 0.78. By contrast, the curve for $x = 1$ exhibits a plateau at intermediate temperatures between two kinks or weak minima indicated by vertical arrows. \label{Slowx}}
\end{figure}

This simple picture ignores that the CEF splitting of the two end members YbRh$_2$Si$_2$ and YbCo$_2$Si$_2$ is rather different. YbRh$_2$Si$_2$ has CEF levels at 0-200-290-500~K~\cite{YRS-06-4}. Scattering from the 3 lower levels and probably even from the full multiplet is held responsible for the thermopower minimum at 80~K~\cite{UK08a}. By contrast, YbCo$_2$Si$_2$ has a much lower overall CEF splitting with levels at 0-46-150-350 K~\cite{Gor00}. Therefore, it is rather surprising that the position of the high-$T$ minimum in $S(T)$ does not change with substitution. It may play a role that both YbRh$_2$Si$_2$ and YbCo$_2$Si$_2$ have a $\Gamma_7$ ground state~\cite{Vya10,YRS-11-5} and that the leading CEF parameter changes smoothly along the series~\cite{YRS-11-4}. These facts are in line with the smooth evolution of the thermopower upon substitution. Nevertheless, in YbCo$_2$Si$_2$ CEF excitations should become relevant at much lower $T$ than in YbRh$_2$Si$_2$. Taking a closer look on the thermopower we find indeed some indication for this: Fig.~\ref{Slowx} shows the thermopower curves for $x = 0.68$, $x = 0.78$, and  $x = 1$ on a larger scale. In addition we plot the corresponding derivatives $\partial S/\partial \log T$. We observe a significant broadening of the high-$T$ thermopower minimum in YbCo$_2$Si$_2$ compared to the samples with mixed Rh/Co content: The high-$T$ minima for $x = 0.68$ and $x = 0.78$ are rather symmetric (on a logarithmic $T$ scale) in contrast to the one for $x = 1$. The corresponding derivatives for $x = 0.68$ and $x = 0.78$ fall almost on top of each other, while the one for pure YbCo$_2$Si$_2$ exhibits a broad plateau between  12~K and 90~K. Most likely the CEF level at 46 K is responsible for the broadening of the thermopower minimum towards lower $T$, while the full multiplet is involved in the minimum at higher $T$.

We would also like to emphasize that the position of the thermopower minimum at elevated $T$ is not directly related to the CEF splitting. To begin with several excited levels contribute to the scattering responsible for the thermopower minimum. The respective effects cannot be simply added as, e.g., for the contributions to the specific heat. Moreover, the exact position of the thermopower minimum depends also on other parameters, in particular the position of the $4f$ level with respect to the Fermi level $\epsilon_{4f}$. The NCA calculations have shown that an increase of $\epsilon_{4f}$ as expected for increasing Co concentration is accompanied by a weak shift of the thermopower minimum towards higher $T$~\cite{TB-05-3}. This may compensate at least partially for the lowering of the overall CEF splitting upon substitution of Rh by Co.

\begin{figure}[tb]
\begin{center}
\includegraphics[width=0.7\textwidth]{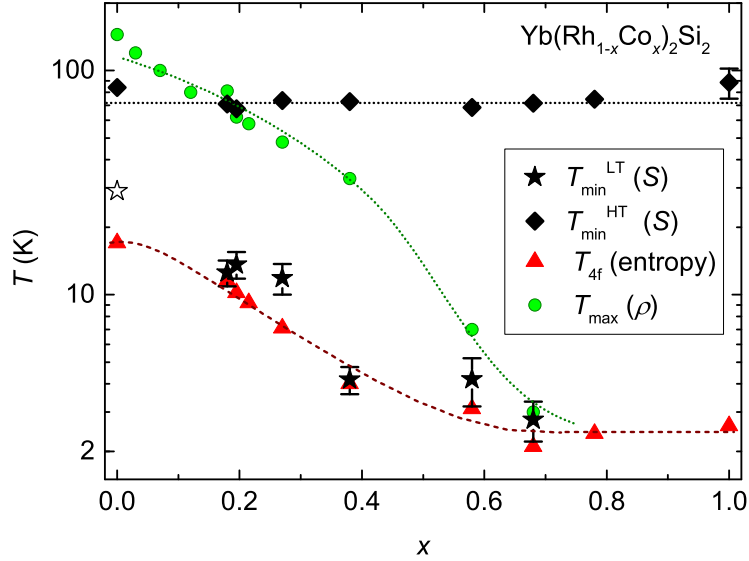}
\end{center}
\caption{(Color online) The positions of the high-$T$ ($T_{\mathrm{min HT}}$) and low-$T$ ($T_{\mathrm{min LT}}$) minima in $S(T)$ for Yb(Rh$_{1-x}$Co$_x$)$_2$Si$_2$ in comparison to the position of the maximum in $\rho(T)$ and to $T_{4f}$ determined from the entropy as explained in the text (both taken from Ref.~\cite{YRS-11-4}). In addition we plot data for $x=0$: $T_{4f}$ taken from Ref.~\cite{Ferstl} and the extrapolated position of the low-$T$ thermopower minimum determined from measurements on Lu$_{1-x}$Yb$_x$Rh$_2$Si$_2$ (open star)~\cite{UK08a}. The lines are guides to the eye. \label{phases}}
\end{figure}

Fig.~\ref{phases} shows the positions of the minima in $S(T)$ at low ($T \mathrm{_{min}^{LT}}$) and high ($T \mathrm{_{min}^{HT}}$) temperature in comparison to other characteristic temperatures taken from Ref.~\cite{YRS-11-4}: the temperature of the maximum in the (total) electrical resistivity $T_{\mathrm{max}}(\rho)$ and the  temperature $T_{4f}$ determined from specific heat measurements as twice the temperature where the magnetic entropy reaches $0.5 R \ln 2$. This temperature is a measure of the magnetic exchange interactions acting on the CEF ground-state doublet and corresponds to $T_K$ in case of dominating Kondo interaction. The value for $T_{4f}$ of YbRh$_2$Si$_2$ is taken from Ref.~\cite{Ferstl}. The error bars for $T \mathrm{_{min}^{LT}}$ and for $T \mathrm{_{min}^{HT}}$ of YbCo$_2$Si$_2$ have been determined as the temperature range for which $S$ deviates less than the scattering of the data from the value at the minimum. The uncertainty of $T \mathrm{_{min}^{HT}}$ of all other compositions is comparable to the symbol size. In addition we also plot the extrapolated position of the low-$T$ thermopower minimum determined for YbRh$_2$Si$_2$ from measurements on Lu$_{1-x}$Yb$_x$Rh$_2$Si$_2$ (open star)~\cite{UK08a}. 

Fig.~\ref{phases} demonstrates again that the low-$T$ minimum in $S(T)$ shifts to lower $T$ with increasing Co concentration, while the position of the high-$T$ minimum is almost independent of $x$. Two additional observations are made: (1) $T \mathrm{_{min}^{LT}}$ roughly follows $T_{4f}$. I.e., the position of the low-$T$ minimum in $S(T)$ is indeed a rough estimate of the Kondo scale in Yb(Rh$_{1-x}$Co$_x$)$_2$Si$_2$. (2) The maximum in the electrical resistivity shifts to lower temperatures with increasing Co content, similar to the low-$T$ minimum in $S(T)$. However, for $x \leq 0.38$ it is observed at much higher $T$. This is at least partly due to the fact that the electrical resistivity contains a contribution from electron-phonon scattering. It is most relevant at elevated temperatures and was not subtracted in Ref.~\cite{YRS-11-4}. An exact evaluation of this component requires accurate knowledge of the sample and contact geometry, which is difficult for small single crystals. For the magnetic contribution to the electrical resistivity we expect the maximum at somewhat lower $T$. Moreover, it appears that the electrical resistivity is less sensitive to CEF excitations in the presence of Kondo scattering than thermopower: While Kondo scattering from ground state and excited CEF levels leads to two separate minima in the thermopower (except for pure YbRh$_2$Si$_2$) only a single maximum is observed in the electrical resistivity. It is due to a combination of Kondo scattering and thermal population of CEF levels and found at some intermediate temperature. For $x \geq 0.58$ a weak hump appears in $\rho(T)$ at $T > T_{\mathrm{max}}(\rho)$ that has been related to the CEF splitting~\cite{YRS-11-4}. For the same concentration range resistivity maximum and low-$T$ thermopower minimum get close to each other. 

\begin{figure}
\begin{center}
\includegraphics[width=0.9\textwidth]{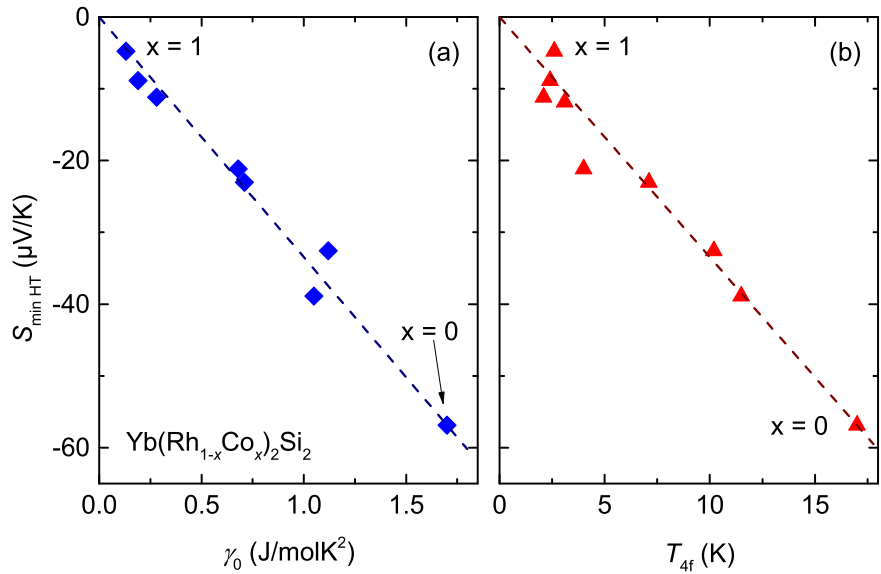}
\end{center}
\caption{(Color online) (a) Value of the thermopower at the high-$T$ minimum, $S \mathrm{_{min}^{HT}}$, vs.~the Sommerfeld coefficient $\gamma_0$ taken from Ref.~\cite{YRS-11-4}. We find a proportionality $S \mathrm{_{min}^{HT}} \propto \gamma _0$. (b) For comparison we also show $S \mathrm{_{min}^{HT}}$, vs.~$T_{4f}$. \label{SvsG}}
\end{figure}

Looking at Fig.~\ref{phases} one might speculate that the maximum in $\rho$ shifts to even lower $T$ and 'passes' the thermopower minimum for higher $x$. However, measurements of $\rho(T)$ performed concomitantly with our thermopower measurements on the sample with $x = 0.78$ (not shown) suggest a maximum in $\rho(T)$ around 2~K, just below our measurement range and close to the anticipated thermopower minimum. Therefore, we suspect that both features remain close to each other also at higher Co concentration. For $x = 1$  magnetic ordering sets in before a maximum in $\rho$ is reached. 

Now we take a look at the lowering of the absolute thermopower values upon increasing Co content $x$. The effect is most obvious around the high-temperature minimum arising from Kondo scattering on the full $4f$ multiplet. This minimum is rapidly suppressed upon substitution. The pure Co-system, YbCo$_2$Si$_2$, exhibits only small absolute thermopower values that are similar in magnitude to those of LuRh$_2$Si$_2$~\cite{UK08a}, the non-magnetic reference to YbRh$_2$Si$_2$. Nevertheless, the minimum around 100 K is still present for YbCo$_2$Si$_2$, which is in line with spectroscopic data revealing a tiny residual hybridization between $4f$ and itinerant states in the material~\cite{YRS-11-4,YRS-14-1}. 
In Fig.~\ref{SvsG}a we plot the value at the high-temperature minimum in $S(T)$, $S \mathrm{_{min}^{HT}}$, vs.~the Sommerfeld coefficient $\gamma_0$ taken from Ref.~\cite{YRS-11-4}. We find a surprisingly good proportionality between these two quantities. For comparison we also show $S \mathrm{_{min}^{HT}}$ vs.~$T_{4f}$ (cf. Fig.~\ref{SvsG}b). In this case the proportionality is less convincing, especially for large $x$. 

Theoretical calculations for the thermopower of Kondo systems with CEF splitting agree to some extend with our observation~\cite{TB-05-3}: In case of an excited level at splitting $\Delta$ above the ground state a maximum (Ce) or minimum (Yb) in the thermopower is predicted at about 0.5 $\Delta / k_{\mathrm{B}}$. The absolute values at the extremum depend sensitively on the parameter choice: Both the hybridization strength and the position of the $4f$ level with respect to the Fermi level, $\epsilon_{4f}$, are relevant, whereat a lowering of the hybridization and an increasing $\epsilon_{4f}$ lead to a lowering of the absolute thermopower over a large parameter range. The situation in Yb(Rh$_{1-x}$Co$_x$)$_2$Si$_2$ is more complicated due to the presence of several excited CEF levels. The hybridization strength may vary for different levels and the respective contributions to the thermopower do not simply add, as mentioned above. Even in case of only one excited CEF level, the clear proportionality between $S(T)$, $S \mathrm{_{min}^{HT}}$ and $\gamma_0$ observed for Yb(Rh$_{1-x}$Co$_x$)$_2$Si$_2$ is not predicted by the calculations and rather surprising. In fact, we compare a quantity determined in the zero-temperature limit and in a magnetically ordered state with a value measured around 90 K. Several points may be relevant for this unexpected behavior: (1) Thermopower contributions from phonon drag and from the diffusion of light charge carriers are most probably small up to 100 K considering the small thermopower of LuRh$_2$Si$_2$. Therefore, the thermopower of the whole series up to $T \mathrm{_{min}^{HT}}$ is dominated by the diffusion contribution from heavy charge carriers. (2) The position of the maximum in $S(T)$ is almost independent of $x$, i.e. we compare thermopower values taken at practically constant $T$. (3) Recent ARPES experiments on YbRh$_2$Si$_2$ revealed a temperature-independent Fermi surface up to about 100 K~\cite{YRS-15-2} and a significant hybridization between $f$ and $d$ states even up to at least 250 K~\cite{YRS-18-1}. In this sense, our thermopower minimum around 90~K is still lying in the low-$T$ regime, and the charge carriers that participate in the transport at 90~K are related to those contributing to $\gamma_0$ taking into account thermal broadening. (4) Moreover, it has been demonstrated recently for a large number of Yb-based Kondo lattices, among them Yb(Rh$_{1-x}$Co$_x$)$_2$Si$_2$, that the deviation of the Yb valence $\nu_{\mathrm{Yb}}$ from 3+ exhibits similar temperature dependencies independent of the respective characteristic ground state Kondo scale~\cite{YRS-18-2}: The $T$ dependence of the change in ($\nu_{\mathrm{Yb}}-3$) normalized to the value of ($\nu_{\mathrm{Yb}}-3$) at 0~K is almost the same for all materials. This observation is so far not understood. However, it means that the change in the Yb valence between 0 K (where $\gamma_0$ is defined) and 90~K (where $S \mathrm{_{min}^{HT}}$ is determined) normalized to the deviation from 3+ at 0~K is the same for all Co concentrations $x$. Considering these points, the direct proportionality between $S \mathrm{_{min}^{HT}}$ and $\gamma_0$ appears at least less amazing. The very special situation realized in our case might also explain, why the proportionality is not found in other $4f$ systems. However, we cannot exclude that its observation for Yb(Rh$_{1-x}$Co$_x$)$_2$Si$_2$ is just an accidental one. Despite this drawback our measurements confirm nicely the expected reduction of the absolute thermopower values upon lowering of the exchange coupling by cobalt substitution, while the persistence of the two weak minima in YbCo$_2$Si$_2$ demonstrates the sensitivity of the thermopower to Kondo scattering.


\section{Summary}

We presented thermopower data on Yb(Rh$_{1-x}$Co$_x$)$_2$Si$_2$. The lowering of $T_K$ with increasing Co concentration gives rise to a low-$T$ minimum in the thermopower, while the absolute thermopower values are rapidly reduced due to the lowering of the exchange coupling between $4f$ and conduction electron states. YbCo$_2$Si$_2$ still exhibits two weak minima in $S(T)$ indicative of a residual hybridization. We find a linear correlation between the thermopower values at the high-$T$ minimum and the Sommerfeld coefficient. The origin of this proportionality is not clear. However it may be related to the recent finding of a rather
stable FS up to temperatures of at least 100 K in YbRh$_2$Si$_2$ and to the similar $T$ dependence of the Yb valence for the whole substitution series.

\begin{acknowledgements}
We acknowledge discussion with K. Kummer, D. V. Vyalikh, and G. Zwicknagel. This work was supported by the German Research foundation (DFG) through project Fermi-NESt. Veljko Zlatic acknowledges the support from the Ministry of Science of Croatia under the bilateral agreement with the USA on scientific and technological cooperation, Project No. 1/2016.  
\end{acknowledgements}


\begin{thebibliography}{spmpsci} 

\bibitem{TB-87-1}
N.E. Bickers, D.L. Cox, J.W. Wilkins, Phys.~Rev.~B \textbf{36}(4), 2036 (1987).
\newblock \url{http://dx.doi.org/10.1103/PhysRevB.36.2036}

\bibitem{TB-97-1}
G.D. Mahan, Phys.~Rev.~B \textbf{56}(18), 11833 (1997).
\newblock \url{http://dx.doi.org/10.1103/PhysRevB.56.11833}

\bibitem{TB-05-3}
V.~Zlati\'{c}, R.~Monnier, Phys.~Rev.~B \textbf{71}, 165109 (2005).
\newblock \url{http://dx.doi.org/10.1103/PhysRevB.71.165109}

\bibitem{TB-76-1}
A.K. Bhattacharjee, B.~Coqblin, Phys.~Rev.~B \textbf{13}(8), 3441 (1976).
\newblock \url{http://dx.doi.org/10.1103/PhysRevB.13.3441}

\bibitem{TB-86-1}
S.~Maekawa, S.~Kashiba, M.~Tachiki, S.~Takahashi, J.~Phys.~Soc.~Jap.
  \textbf{55}(9), 3194 (1986).
\newblock \url{http://dx.doi.org/10.1143/JPSJ.55.3194}

\bibitem{US12a}
A.P. Pikul, U.~Stockert, A.~Steppke, T.~Cichorek, S.~Hartmann,
  N.~Caroca-Canales, N.~Oeschler, M.~Brando, C.~Geibel, F.~Steglich,
  Phys.~Rev.~Lett. \textbf{108}, 066405 (2012).
\newblock \url{http://doi.org/10.1103/PhysRevLett.108.066405}

\bibitem{TB-01-3}
D.~Huo, J.~Sakurai, O.~Maruyama, T.~Kuwai, Y.~Isikawa, J.~Magn.~Magn.~Mater.
  \textbf{226-230}, 202 (2001).
\newblock \url{https://doi.org/10.1016/S0304-8853(01)00105-6}

\bibitem{TB-99-3}
D.~Andreica, K.~Alami-Yadri, D.~Jaccard, A.~Amato, A.~Schenck, Physica B
  \textbf{259-261}, 144 (1999).
\newblock \url{https://doi.org/10.1016/S0921-4526(98)00832-1}

\bibitem{YRS-00-1}
O.~Trovarelli, C.~Geibel, S.~Mederle, C.~Langhammer, F.M. Grosche,
  P.~Gegenwart, M.~Lang, G.~Sparn, F.~Steglich, Phys.~Rev.~Lett.
  \textbf{85}(3), 626 (2000).
\newblock \url{http://doi.org/10.1103/PhysRevLett.85.626}

\bibitem{YRS-02-2}
P.~Gegenwart, J.~Custers, C.~Geibel, K.~Neumaier, T.~Tayama, K.~Tenya,
  O.~Trovarelli, F.~Steglich, Phys.~Rev.~Lett. \textbf{89}(5), 056402 (2002).
\newblock \url{http://doi.org/10.1103/PhysRevLett.89.056402}

\bibitem{YRS-03-2}
J.~Custers, P.~Gegenwart, H.~Wilhelm, K.~Neumaier, Y.~Tokiwa, O.~Trovarelli,
  C.~Geibel, F.~Steglich, C.~P\'{e}pin, P.~Coleman, Nature \textbf{424}(6948),
  524 (2003).
\newblock \url{http://doi.org/ 10.1038/nature01774}

\bibitem{YRS-06-8}
P.~Gegenwart, Y.~Tokiwa, T.~Westerkamp, F.~Weickert, J.~Custers, J.~Ferstl,
  C.~Krellner, C.~Geibel, P.~Kerschl, K.H. M\"{u}ller, F.~Steglich, New Journal
  of Physics \textbf{8}, 171 (2006).
\newblock \url{http://doi.org/10.1088/1367-2630/8/9/171}

\bibitem{UK08a}
U.~K\"ohler, N.~Oeschler, F.~Steglich, S.~Maquilon, Z.~Fisk, Phys.~Rev.~B
  \textbf{77}, 104412 (2008).
\newblock \url{http://dx.doi.org/10.1103/PhysRevB.77.104412}

\bibitem{YRS-11-1}
K.~Kummer, Y.~Kucherenko, S.~Danzenb\"{a}cher, C.~Krellner, C.~Geibel, M.G.
  Holder, L.V. Bekenov, T.~Muro, Y.~Kato, T.~Kinoshita, S.~Huotari,
  L.~Simonelli, S.L. Molodtsov, C.~Laubschat, D.V. Vyalikh, Phys.~Rev.~B
  \textbf{84}, 245114 (2011).
\newblock \url{http://dx.doi.org/10.1103/PhysRevB.84.245114}

\bibitem{YRS-06-4}
O.~Stockert, M.M. Koza, J.~Ferstl, A.P. Murani, C.~Geibel, F.~Steglich, Physica
  B \textbf{378-380}, 157 (2006).
\newblock \url{http://doi.org/10.1016/j.physb.2006.01.059}

\bibitem{YRS-11-5}
C.~Klingner, C.~Krellner, M.~Brando, C.~Geibel, F.~Steglich, New~J.~Phys.
  \textbf{13}, 083024 (2011).
\newblock \url{http://doi.org/10.1088/1367-2630/13/8/083024}

\bibitem{YRS-11-4}
C.~Klingner, C.~Krellner, M.~Brando, C.~Geibel, F.~Steglich, D.V. Vyalikh,
  K.~Kummer, S.~Danzenb\"{a}cher, S.L. Molodtsov, C.~Laubschat, T.~Kinoshita,
  Y.~Kato, T.~Muro, Phys.~Rev.~B \textbf{83}, 144405 (2011).
\newblock \url{http://dx.doi.org/10.1103/PhysRevB.83.144405}

\bibitem{Gor00}
E.A. Goremychkin, R.~Osborn, J.~Appl.~Phys. \textbf{87}, 6818 (2000).
\newblock \url{https://doi.org/10.1063/1.372852}

\bibitem{YRS-14-1}
M.~G\"{u}ttler, K.~Kummer, S.~Patil, M.~H\"{o}ppner, A.~Hannaske,
  S.~Danzenb\"{a}cher, M.~Shi, M.~Radovic, E.~Rienks, C.~Laubschat, C.~Geibel,
  D.V. Vyalikh, Phys.~Rev.~B \textbf{90}, 195138 (2014).
\newblock \url{http://dx.doi.org/10.1103/PhysRevB.90.195138}

\bibitem{YRS-08-3}
T.~Westerkamp, P.~Gegenwart, C.~Krellner, C.~Geibel, F.~Steglich, Physica B
  \textbf{403}, 1236 (2008).
\newblock \url{https://doi.org/10.1016/j.physb.2007.10.114}

\bibitem{YRS-13-1}
S.~Lausberg, A.~Hannaske, A.~Steppke, L.~Steinke, T.~Gruner, L.~Pedrero,
  C.~Krellner, C.~Klingner, M.~Brando, C.~Geibel, F.~Steglich, Phys.~Rev.~Lett.
  \textbf{110}, 256402 (2013).
\newblock \url{http://doi.org/10.1103/PhysRevLett.110.256402}

\bibitem{YRS-06-5}
S.~Hartmann, U.~K\"{o}hler, N.~Oeschler, S.~Paschen, C.~Krellner, C.~Geibel,
  F.~Steglich, Physica B \textbf{378-380}, 70 (2006).
\newblock \url{http://doi.org/10.1016/j.physb.2006.01.028}

\bibitem{Vya10}
D.V. Vyalikh, S.~Danzenb\"acher, Y.~Kucherenko, K.~Kummer, C.~Krellner,
  C.~Geibel, M.G. Holder, T.K. Kim, C.~Laubschat, M.~Shi, L.~Patthey,
  R.~Follath, S.L. Molodtsov, Phys.~Rev.~Lett. \textbf{105}, 237601 (2010).
\newblock \url{http://doi.org/10.1103/PhysRevLett.105.237601}

\bibitem{Ferstl}
J.~Ferstl,  (PhD thesis, Techn.~Univ. of Dresden, 2007)

\bibitem{YRS-15-2}
K.~Kummer, S.~Patil, A.~Chikina, M.~G\"{u}ttler, M.~H\"{o}ppner, A.~Generalov,
  S.~Danzenb\"{a}cher, S.~Seiro, A.~Hannaske, C.~Krellner, Y.~Kucherenko,
  M.~Shi, M.~Radovic, E.~Rienks, G.~Zwicknagl, K.~Matho, J.~Allen,
  C.~Laubschat, C.~Geibel, D.V. Vyalikh, Phys.~Rev.~X \textbf{5}, 011028
  (2015).
\newblock \url{https://doi.org/10.1103/PhysRevX.5.011028}

\bibitem{YRS-18-1}
D.~Leuenberger, J.A. Sobota, S.L. Yang, H.~Pfau, D.J. Kim, S.K. Mo, Z.~Fisk,
  P.S. Kirchmann, Z.X. Shen, Phys.~Rev.~B \textbf{97}, 165108 (2018).
\newblock \url{http://dx.doi.org/10.1103/PhysRevB.97.165108}

\bibitem{YRS-18-2}
K.~Kummer, C.~Geibel, C.~Krellner, G.~Zwicknagel, C.~Laubschat, N.B. Brookes,
  D.V. Vyalikh, Nature~Commun. \textbf{9}, 2011 (2018).
\newblock \url{http://dx.doi.org/10.1038/s41467-018-04438-8}

\end{thebibliography}
\end{document}